\documentclass[twocolumn,superscriptaddress,preprintnumbers,amsmath,amssymb,longbibliography]{revtex4-1}
\usepackage{graphicx}
\usepackage{dcolumn}
\usepackage{bm}
\usepackage{color}

\usepackage{dcolumn}

\makeatletter
\@ifundefined{textcolor}{}
{%
 \definecolor{BLACK}{gray}{0}
 \definecolor{WHITE}{gray}{1}
 \definecolor{RED}{rgb}{1,0,0}
 \definecolor{GREEN}{rgb}{0,1,0}
 \definecolor{BLUE}{rgb}{0,0,1}
 \definecolor{CYAN}{cmyk}{1,0,0,0}
 \definecolor{MAGENTA}{cmyk}{0,1,0,0}
 \definecolor{YELLOW}{cmyk}{0,0,1,0}
}

\begin{document}

\title{Strong Photon Blockade Mediated by Optical Stark Shift in a Single-Atom-Cavity System}

\author{Jing Tang}
\affiliation{Laboratory of Quantum Engineering and Quantum Metrology, School of Physics and Astronomy, Sun Yat-Sen University (Zhuhai Campus), Zhuhai 519082, People's Republic of China}

\author{Yuangang Deng}
\email{dengyg3@mail.sysu.edu.cn}
\affiliation{Laboratory of Quantum Engineering and Quantum Metrology, School of Physics and Astronomy, Sun Yat-Sen University (Zhuhai Campus), Zhuhai 519082, People's Republic of China}

\author{Chaohong Lee}
\email{lichaoh2@mail.sysu.edu.cn}
\affiliation{Laboratory of Quantum Engineering and Quantum Metrology, School of Physics and Astronomy, Sun Yat-Sen University (Zhuhai Campus), Zhuhai 519082, People's Republic of China}
\affiliation{State Key Laboratory of Optoelectronic Materials and Technologies, Sun Yat-Sen University (Guangzhou Campus), Guangzhou 510275, People's Republic of China}
\affiliation{Synergetic Innovation Center for Quantum Effects and Applications, Hunan Normal University, Changsha 410081, People's Republic of China}

\date{\today}

\begin{abstract}
We propose a theoretical scheme to achieve strong photon blockade via a single atom in cavity. By utilizing optical Stark shift, the dressed-state splitting between higher and lower branches is enhanced, which results in significant increasing for lower (higher) branch and decreasing for higher (lower) branch at the negative (positive) Stark shift, and dominates the time-evolution of photon number oscillations as well. Furthermore, the two-photon excitation is suppressed via quantum interference under optimal phase and Rabi frequency of an external microwave field. It is shown that the interplay between the quantum interference and the enhanced vacuum-Rabi splitting gives rise to a strong photon blockade for realizing high-quality single photon source beyond strong atom-cavity coupling regime. In particular,  by tuning the optical Stark shift, the second-order correlation function for our scheme has three orders of magnitude smaller than Jaynes-Cummings model and correspondingly there appears a large cavity photon number as well. Our proposal may implicate exciting opportunities for potential applications on quantum network and quantum information processing.
\end{abstract}

\maketitle
\section{INTRODUCTION}

The realization and manipulation of nonclassical light fields at single-photon level plays an essential role in quantum information science~\cite{Duan01, Duan04, Pieter07, Kimble08, Giovannetti2011}. It offers unprecedented scientific opportunities for exploring various applications in quantum computing~\cite{Bennett00, Knill01, Buluta11}, quantum key distribution~\cite{Scarani09, Brien09}, quantum simulation~\cite{Buluta108, Georgescu14}, and quantum metrology~\cite{Giovannetti2011}. Recently, the controllable single-photon emitters have been extensively explored in different quantum systems, including atom-cavity coupled systems~\cite{McKeever04, Wilk07, Le16},  quantum-dot-cavity coupled systems~\cite{Hennessy07, Toishi09, Kai15}, coupled optomechanical systems~\cite{Rabl11,Nunnenkamp11,Wang15}, coupled optical-fiber systems~\cite{Xu07, Xu08}, waveguide-QED systems\cite{Mirza16,  Mirza17, Gheeraert18, Zheng11, Mirza2016}, and circuit-QED systems~\cite{Hoffman11,Lang11, Liu14}, in which the quantum statistics of the output photons show a sub-Poissonian distribution due to the photon blockade (PB)~\cite{Zou90}.

There are two typical PB mechanisms that can be used for generating a single photon in cavity QED. One is based on strong coupling with a large vacuum Rabi splitting in a cavity with a single two-level atom~\cite{Imamoglu97,  Hartmann07, Chang14}, which gives rise to an anharmonic ladder of dressed states in the energy spectrum. When the energy gap of the two-photon resonance is larger than the cavity decay rate, a first photon generated in the cavity will block the transmission of a second photon~\cite{Imamoglu97}. This mechanism relying on strong energy-spectrum anharmonicity, is referred to as conventional PB, and remains a great challenge, despite some experimental advances in realizing strong coupling in a high-finesse cavity~\cite{Birnbaum2005, Hartmann06, Fink08, Faraon08, Dayan08, Chang08, Brossard10, xu18, xu2018}.

\begin{figure}[ht]
\includegraphics[width=1\columnwidth]{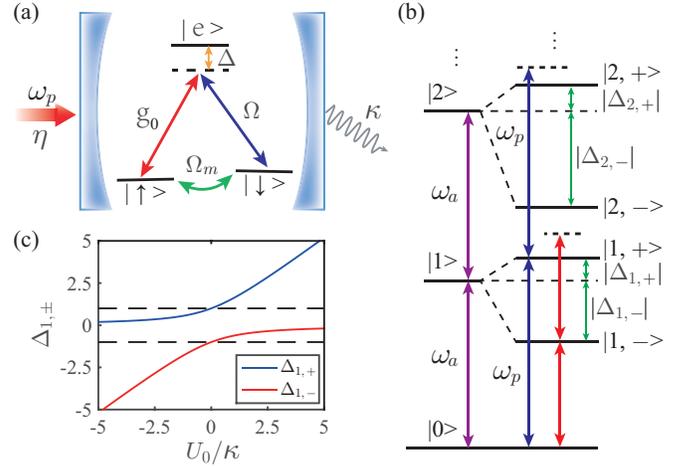}
\caption{(a) Schematic illustration of creation of PB in a cavity with a single three-level atom. In order to realize quantum interference, a microwave field is employed to couple the spin states $|\uparrow\rangle$ and $|\downarrow\rangle$. (b) Typical anharmonic energy spectrum with ignorable weak driving and microwave fields. Here, $\Delta_{n,\pm}$ is the energy splitting for the n-$th$ excited state, defined by Eq.~(\ref{gap}). (c) Vacuum-Rabi splitting [$\Delta_{1, -}$(red line) and $\Delta_{1, +}$ (blue line)] versus optical Stark shift $U_0$  for  $g/\kappa=1$. The horizontal dashed lines at $\Delta_{1, \pm}=\pm1$ are guides to the eye.} \label{scheme}
\end{figure}%

In contrast to conventional PB, unconventional PB has led to tremendous advances in achieving strong antibunching of photons by using quantum interference~\cite{Liew10, Bamba2011, Majumdar2012, Tang15, Zhou15, Flayac17,  Sarma18, Snijders18,Vaneph18}, in which the basic principle is based on constructive interference between different quantum transition paths from the ground state to a two-photon excitation state. Unconventional PB has been observed experimentally in systems ranging from a single-quantum-dot cavity~\cite{Snijders18} to, currently, two coupled superconducting resonators~\cite{Vaneph18}. However, the quantum interference mechanism cannot eliminate all multiphoton excitations simultaneously, even if the two-photon excitation state is suppressed completely. In fact, we find that the strong PB based on quantum interference also requires sufficient energy-spectrum anharmonicity. Therefore, it is of great interest to answer whether there exists a new system that utilizes advantages of both conventional and unconventional PB to realize an ideal single-photon source beyond strong coupling regime. An affirmative answer will significantly enhance our understanding of fundamental quantum phenomena in cavity QED and stimulate the relevant applications in quantum information science.

In this paper, we study such a scenario in which strong PB is generated via Raman and microwave fields with a single atom in an optical cavity. We show that, depending on the strength and sign of the optical Stark shift, the vacuum Rabi splitting of the system in the red or blue sideband can be largely enhanced, and this also determines the photon-number oscillations in the time evolution. Remarkably, strong photon antibunching with a relatively large cavity photon number is predicted when an enhanced vacuum Rabi splitting with generation of a strong energy-spectrum anharmonicity is combined with quantum interference with elimination of two-photon excitation beyond the strong-coupling regime. The proposed scheme has the advantage that it does not require strong atom-cavity coupling, which may facilitate the study of quantum networks and high-quality multiphoton sources. In addition, the excitation-energy spectrum is sensitive to the sign of the optical Stark shift, which could provide a new method for diagnosis of atomic states by means of quantum nondemolition measurement of the statistical properties of photons.

\section{MODEL AND ENERGY SPECTRUM}

We consider a single three-level atom inside an optical cavity. The atom may occupy one excited state $|e\rangle$ and two hyperfine ground states $|\uparrow\rangle$ and $|\downarrow\rangle$~\cite{Mucke10, Tobias10}. A bias magnetic field ${\bf B}$ is applied along the cavity axis, which produces a large enough Zeeman shift $\hbar\omega_Z$ between the two hyperfine ground states and defines the quantization $z$ axis. As shown in Fig. \ref{scheme}(a), the magnetic quantum numbers of the electronic states satisfy $m_{\uparrow}=m_{\downarrow} +1$ and  $m_{|e\rangle}=m_{\downarrow}$. Specifically, the single atom is pumped by a $\pi$-polarized classical laser field with frequency $\omega_L$ propagating orthogonal to the cavity axis, which drives an atomic transition $|\downarrow\rangle\leftrightarrow |e\rangle$ with Rabi frequency $\Omega$ and atom-pump detuning $\Delta$. As to the cavity, Bragg scattering into the $\pi$-polarized cavity mode (with polarization along the cavity axis) is forbidden in the configuration of the laser and cavity used [Fig. \ref{scheme}(a)]. Therefore the cavity only supports only a $\sigma$-polarized light field originating from the vacuum-stimulated Raman transition\cite{Deng14}. Then the atomic transition $|\uparrow\rangle\leftrightarrow|e\rangle$ is driven by the  $\sigma$-polarized cavity field with a single-photon coupling strength  $g_{0}$, where the bare cavity frequency and decay rate are $\omega_{c}$ and $\kappa$, respectively. In addition, the cavity is also driven by a weak laser field with frequency $\omega_p$ and amplitude $\eta$.

In the limit of large atom-pump detuning, $|g_0/\Delta|\ll 1$ and $|\Omega/\Delta|\ll 1$, the excited state $|e\rangle$ can be adiabatically eliminated. Using the rotating-wave approximation, the relevant Hamiltonian of the single-two-level-atom-cavity system is given by
\begin{align}\label{Hamiltonian}
\hat{H}/\hbar&= \Delta_c \hat{a}^\dag\hat{a} + (U_0 \hat{a}^\dag\hat{a} -\Delta_a) |\uparrow\rangle\langle\uparrow|+\eta(\hat{a}^\dag  + \hat{a}) \nonumber \\
&+[(g \hat{a}^\dag + \Omega_me^{i\theta}) |\uparrow\rangle\langle\downarrow|+ {\rm H.c.}],
\end{align}
where $\hat{a}^\dag$ ($\hat{a}$) is the creation (annihilation) operator of the single cavity mode, $g=-\Omega g_0/\Delta$ is the Raman coupling strength, $U_0=-g_0^2/\Delta$ is the optical Stark shift, $\Delta_c=\omega_c-\omega_p$ is the cavity-light detuning, and $\Delta_a=\omega_Z+\omega_L-\omega_p-\Omega^2/\Delta$ is the effective two-photon detuning. In addition, the two ground states are also coupled by a microwave field at the Rabi frequency $\Omega_m e^{i\theta}$. Here $\theta$ is the tunable phase difference between the Raman coupling and the microwave field. As we shall see below, the additional microwave field plays an essential role in generating a large cavity photon number and strong PB based on the quantum interference mechanism.

In the limit of weak driving and microwave fields, $|\eta/g|\ll 1$ and $|\Omega_{m}/g|\ll 1$, the total excitation number is conserved.
Therefore the relevant Hilbert space of the system can be restricted to $|n,\uparrow \rangle$ and $|n-1,\downarrow \rangle$, where $n$ represents the number of photons in the cavity. Then, the Hamiltonian in Eq.~(\ref{Hamiltonian}) can be diagonalized, corresponding to the energy eigenvalues $E_{n\pm}=n \Delta_c + n U_0/2 \pm\sqrt{n^2U_0^2+4ng^2}/2$ when we set $\Delta_a=\Delta_c$.
Here the energy splittings of the n-$th$  pair of dressed states $|n,\pm\rangle$ are given by
\begin{align}\label{gap}
\Delta_{n,\pm} =[nU_0 \pm\sqrt{n^2U_0^2+4ng^2}]/2,
\end{align}
where $+$ ($-$) denotes the higher (lower) branch. It is clear that the splittings ($\sqrt{n^2U_0^2+4ng^2}$) between the higher and lower branches of the dressed states are enhanced (roughly linearly increased) when the optical Stark shift $|U_0|$ and the excitation number $n$ are increased. In Fig.~\ref{scheme}(b), we show the anharmonicity ladder of the energy spectrum. Interestingly, an asymmetry in the dressed-state splitting between red and blue light-cavity detunings is observed for a nonzero optical Stark shift $U_0$.

Remarkably, $U_0$ plays an important role in modifying the vacuum Rabi splitting $|\Delta_{1,\pm}|$, which significantly increases (decreases) the values of $|\Delta_{1,-}|$ ($|\Delta_{1,+}|$) for the lower (higher) branch when $U_0<0$ ($U_0>0$). The variation in the vacuum Rabi splitting $\Delta_{1,\pm}$ as a function of $U_0$ with $g/\kappa=1$ is displayed in Fig.~\ref{scheme}(c). As can be seen, $U_0$ gives rise to an asymmetric vacuum Rabi splitting between red and blue light-cavity detunings. For red detuning of the excitation, $\Delta_{1, -}$ gradually approaches to zero with increasing positive $U_0$. But for negative $U_0$, $|\Delta_{1, -}|$ is clearly dependent on the value of $U_0$, with approximately linear growth as $|U_0|$ increases. In addition, similar behavior of $\Delta_{1, +}$ is observed for blue detuning of the excitation.

In conventional PB, the blockade mechanism is ascribed to a reinforced vacuum Rabi splitting, which  generates strong energy-spectrum anharmonicity. When one photon is absorbed resonantly with the higher or lower branch, one photon excitation will "blocks" a second photon excitation due to the larger two-excitation energy gap, leading to an orderly output of photons one by one with strong photon antibunching. As expected, the enhanced vacuum Rabi splitting (energy-spectrum anharmonicity) induced by $U_0$ facilitates the generation of strong PB due to a strong reduction of the spectral overlap with higher-excitation states [Fig.~\ref{scheme}(b)]. Recently, enhanced PB has been experimentally demonstrated via detuning the quantum-dot and cavity resonances\cite{Kai15}, for which the mechanism is similar to that based on the optical Stark shift with tuning of the vacuum Rabi splitting. However, the  optical Stark shift will further induce strong asymmetry in the higher manifolds of the dressed states in our proposal.

\section{QUANTUM INTERFERENCE}

Unlike conventional PB, which requires strong atom-cavity coupling, unconventional PB can be achieved by utilizing quantum interference even in the weak-coupling regime~\cite{Tang15} through constructing two different quantum transition paths between the cavity field and the classical pumping field. In the analogous configuration of Ref.~\cite{Tang15} for a single-quantum-dot-cavity system, the present model of Eq.~(\ref{Hamiltonian}) with application of a Raman field and microwave fields should also possess the property that quantum interference can suppress two-photon excitation. Therefore, strong PB will emerge when quantum interference and a large vacuum Rabi splitting induced by an additional optical Stark shift are combined. To proceed further, we calculate the optimal conditions for quantum interference in the system in the following.

In the presence of a microwave field and a weak driving field, the excitation number $n$ is not a good quantum number, which indicates that an exact analytical solution for the system does not exist. Nevertheless, the Hamiltonian~(\ref{Hamiltonian}) can be approximatly diagonalized in subspaces defined by a large enough truncation excitation number of the system. Without loss of generality, the wave function of the system is formally given by~\cite{Tang15}
\begin{align}
|\psi\rangle &= \sum_{n=0}^{\infty} C_{n,\uparrow} |n, \uparrow\rangle
+\sum_{n=1}^{\infty} C_{n-1, \downarrow}|n-1,\downarrow\rangle,
\end{align}
where $|C_{n,\uparrow}|^2$ and $|C_{n-1,\downarrow}|^2$ denote the occupation probability for eigenstates $|n, \uparrow\rangle$ and $|n-1, \downarrow\rangle$, respectively. In the single-PB regime, we need only to truncate the two-photon ($n=2$) excitation subspaces. As a result, the wave function for the system can be approximated as
\begin{align}
|\psi\rangle  &=
\sum_{\sigma=\uparrow,\downarrow}\left[C_{0,\sigma}|0,\sigma\rangle + C_{1,\sigma}|1,\sigma\rangle \right]+C_{2,\uparrow}|2,\uparrow\rangle,
\end{align}
By substituting $|\psi\rangle$ into the Schr\"{o}dinger equation, the probability coefficients in the steady-state solutions are determined from the following equations:
\begin{align}
&\eta C_{1,\uparrow} \!+ \Omega_m e^{i\theta} C_{0,\downarrow} \!= 0, \nonumber \\
&\eta C_{1,\uparrow} \!+ \sqrt{2} \Delta_1 C_{2,\uparrow} \!+g C_{1,\downarrow} \!= 0, \nonumber \\
&\Omega_m e^{-i\theta} C_{1,\uparrow} \!+ \eta C_{0,\downarrow} \!+  \sqrt{2}g C_{2,\uparrow} \!+\Delta_2 C_{1,\downarrow} \!= 0,\nonumber \\
&\Omega_m e^{-i\theta} C_{0,\uparrow} \!+ g C_{1,\uparrow} \!+ (\Delta_c -i\gamma_g) C_{0,\downarrow} \!+\eta C_{1,\downarrow} \!=0, \nonumber \\
&\eta C_{0,\uparrow} \!+ \Delta_1 C_{1,\uparrow} \!+ g C_{0,\downarrow} \!+\sqrt{2} \eta C_{2,\uparrow} \!+\Omega_m e^{i\theta} C_{1,\downarrow} \!=\! 0,\label{stead-equation}
\end{align}
where $\gamma_g$ is the loss rate of the atomic ground state, and $\Delta_1=\Delta_c +U_0-i\kappa$ and $\Delta_2=2\Delta_c  -i\kappa-i\gamma_g$ are introduced as shorthand notation.

\begin{figure}[ht]
\includegraphics[width=0.65\columnwidth]{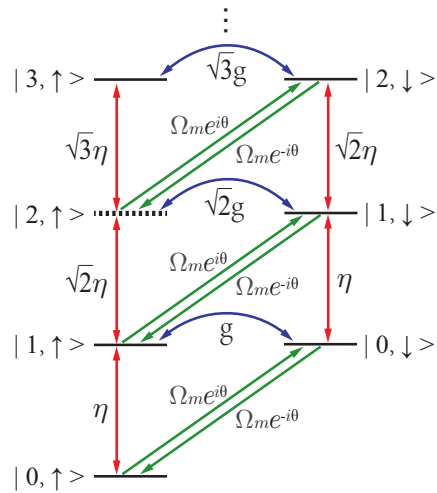}
\caption{Quantum transition paths in the atom-cavity system~(\ref{Hamiltonian}). The two-photon excited state $|2,\uparrow\rangle$ (dashed line) is eliminated by employing destructive quantum interference.} \label{path}
\end{figure}%

For the single-photon-excitation limit, multiphoton excitation with $n\geq2$ must be suppressed. At least a two-photon excitation probability $C_{2,\uparrow}=0$ is required first. For a nonzero microwave-field coupling, the nontrivial solution of Eqs.~(\ref{stead-equation}) is given by
\begin{align}\label{OEQ}
\theta_{\rm{opt}} &={\arctan} (\frac{\kappa+\gamma_g}{2\Delta_c}),\nonumber \\
\Omega_{m,{\rm opt}} &= \frac{{\eta}(\mathcal {R} + \sqrt{\mathcal
{R}^2 +4})}{2},
\end{align}
where ${\mathcal R} ={\sqrt{4{\Delta^2_c}+{(\kappa+\gamma_g)}^2}}/{g}$ and $0\leq\theta_{\rm{opt}}/\pi\leq 1$. The solutions of Eq.~(\ref{OEQ}) for the optimal phase $\theta_{\rm opt}$ and Rabi frequency $\Omega_{m,{\rm opt}}$ denote the conditions for eliminating two-photon excitation by quantum interference. To generate a large cavity output, we must focus on single-photon resonance excitation with a fixed $\Delta_c=-\Delta_{1,\pm}$ in Eq.~(\ref{OEQ}). Therefore, the optimal parameters are significantly dependent on the optical Stark shift $U_0$.

The essential mechanism for generating $C_{2,\uparrow}=0$ is ascribed to destructive quantum interference between different quantum transition paths even for a weak coupling strength of the system, see Fig.~\ref{path}. This interference can occur, for example, between the transitions $\left|1,\uparrow\right\rangle \xrightarrow{\sqrt{2} \eta} \left|2,\uparrow\right\rangle$ induced by the driving field and $\left|1,\uparrow\right\rangle \xrightarrow{\Omega_m e^{i\theta}} \left|1,\downarrow\right\rangle \xrightarrow{\sqrt{2} g} \left|2,\uparrow\right\rangle$ induced by the combination of both the microwave field and the cavity field, after the state $|1,\uparrow\rangle$ is excited from the initial state $|0,\uparrow\rangle$ by the driving field.
Then the probability of two-photon excitation is zero when the optimal conditions of Eq.~(\ref{OEQ}) are satisfied. We should note that the quantum interference mechanism ($C_{2,\uparrow}=0$) vanishes in the absence of a microwave field ($\Omega_m=0$).


\section{NUMERICAL RESULTS}

To further demonstrate PB, we calculate the quantum statistical properties of the system by solving the quantum master equation using the Quantum Optics Toolbox~\cite{Tan99}. Including the dissipation and pure dephasing of the system, the Liouvillian superoperator ${\cal{L}}$ of the atom-cavity coupled system satisfies
\begin{equation}\label{master equation}%
{\cal{L}}\rho = -i [\hat{H}, {\rho}] + \frac{\kappa}{2} \mathcal
{\cal{D}}[\hat{a}]\rho + \frac{\gamma_g}{2} \mathcal
{\cal{D}}[\hat{\sigma}_{-}]\rho + \frac{\gamma_d}{2}\mathcal{\cal{D}}[\hat{\sigma}_{+}\hat{\sigma}_{-}]\rho ,
\end{equation}
where $\rho$ is the density matrix of the atom-cavity system,  $\gamma_d$ is the pure dephasing, $\sigma_-=|\uparrow\rangle \langle \downarrow|$ is the lowering spin-$1/2$ operator, and $\mathcal {D}[\hat{o}]\rho=
2\hat{o} {\rho} \hat{o}^\dag - \hat{o}^\dag \hat{o}{\rho} - {\rho}
\hat{o}^\dag \hat{o}$ denotes the Lindblad type of dissipation. In experiments, the second-order correlation function is a key physical quantity for PB, and is defined as~\cite{Glauber63}
\begin{align} \label{inequation}%
g^{(2)}(\tau)=\frac{<\hat{a}^\dagger(t)\hat{a}^\dagger(t+\tau)\hat{a}(t+\tau)\hat{a}(t)>}{<\hat{a}^\dagger(t)\hat{a}(t)><\hat{a}^\dagger(t+\tau)\hat{a}(t+\tau)>}.
\end{align}
For the steady-state solution with ${\cal L}\rho_s = 0$ \cite{Tan99}, the second-order correlation function for zero time delay, $g^{(2)}(0)={\rm Tr}(\hat{a}^\dagger\hat{a}^\dagger
\hat{a}\hat{a}\rho_s)/{n}^2_s$, and the steady-state intracavity photon number, ${n}_s = {\rm Tr}(\hat{a}^\dag
\hat{a}\rho_s)$, are directly calculated by solving the quantum master equation. The nonclassical
antibunching of photons is characterized by $g^{(2)}(0) < g^{(2)}(\tau)$, and $g^{(2)}(0)=0$ is a signature of ideal PB \cite{Mandel1995}.

Specifically, we consider a single ultracold spin-$\frac{1}{2}$ atom in a high-finesse optical cavity with a cavity decay rate $\kappa=2\pi\times 147$ kHz~\cite{Li2017}. The intracavity photon number for an empty cavity is $n_0=(\eta/\kappa)^2=0.02$~\cite{Birnbaum2005}, the atomic decay rate for the ground state is $\gamma_g/\kappa=10^{-3}$ and the pure dephasing is set to $\gamma_d/\gamma_g=1.0$. Taking the cavity decay rate $\kappa$ as the energy unit, the theoretically tunable parameters include the atom-cavity (microwave) coupling strength $g$ ($\Omega_m$), the phase difference $\theta$, the optical Stark shift $U_0$, and the cavity-light detuning $\Delta_c$. Note that $\Omega_m$ and $\theta$ can be fixed by applying the quantum interference conditions of Eq.~(\ref{OEQ}). In the following, we focus only on the red-sideband excitation in Eq.~(\ref{OEQ}). In fact, we have checked that the results for blue-sideband excitation display similar behavior. As a result, the free parameters in our system are reduced to $g$, $U_0$, and $\Delta_c$.

\begin{figure}[ht]
\includegraphics[width=1\columnwidth]{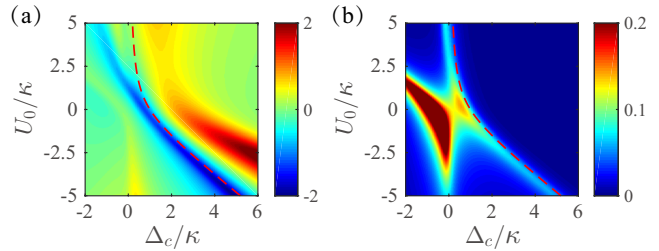}
\caption{(a) Second-order correlation function and (b) intracavity photon number as a function of cavity-light detuning $\Delta_c$ and optical Stark shift $U_0$ for the optimal parameters $\theta_{\rm{opt}}$ and $\Omega_{m,\rm{opt}}$ with $g/\kappa=1$. The red dashed line shows the vacuum Rabi splitting ($|\Delta_{1,-}|$) of the system in the red sideband. The colors with blue-red gradient shading indicate the values of log[$g^{(2)}(0)$] in (a) and ${n}_s$ in (b).} \label{main}
\end{figure}%

Figure~\ref{main}(a) displays our numerical results for $g^{(2)}(0)$ in the $\Delta_c$-$U_0$ parameter plane under the optimal quantum interference conditions of Eq.~(\ref{OEQ}) with $g/\kappa=1$. The red dashed line shows the red-sideband vacuum Rabi splitting.
As expected, $g^{(2)}(0)$ obviously depends on the sign of the optical Stark shift $U_0$ and the cavity-light detuning $\Delta_c$; the asymmetric features and the optimal $g^{(2)}(0)$ [the blue regime in Fig.~\ref{main}(a)] are consistent with the analytic energy spectrum, especially for negative $U_0$, in spite of the driving fields and the cavity decay are being included. Fig.~\ref{main}(b) shows the mean cavity photon number $n_s$ with the same parameters as in Fig.~\ref{main}(a). As can be seen, ${n}_s$ also displays an asymmetric behavior dependent on the sign of $U_0$. When $\Delta_c$ is tuned, ${n}_s$ exhibits two peaks near the red and blue sidebands of the single-photon resonance points. Interestingly, the cavity field simultaneously hosts strong PB with $g^{(2)}(0)<0.01$ and a large cavity output in the red sideband with  $\Delta_c\approx\Delta_{1,-}$ (red dashed line), which is in good agreement with the vacuum Rabi splitting in Eq.~(\ref{gap}).

We emphasize that the strong antibunching is contributed by the use of quantum interference for suppressing two-photon excitation and the optical-Stark-shift-enhanced vacuum Rabi splitting with increasing energy-spectrum anharmonicity. As shown in Fig.~\ref{path}, multiphoton excitations ($n\geq3$) also need to be suppressed simultaneously to generate strong PB, which indicates that the quantum interference mechanism is  not sufficient to guarantee the realization of strong PB, as we will discuss below. Nevertheless, the multiphoton excitations are far off resonance due to the strong anharmonicity of the energy spectrum enhanced by the optical Stark shift in our system.

\begin{figure}[ht]
\includegraphics[width=1.05\columnwidth]{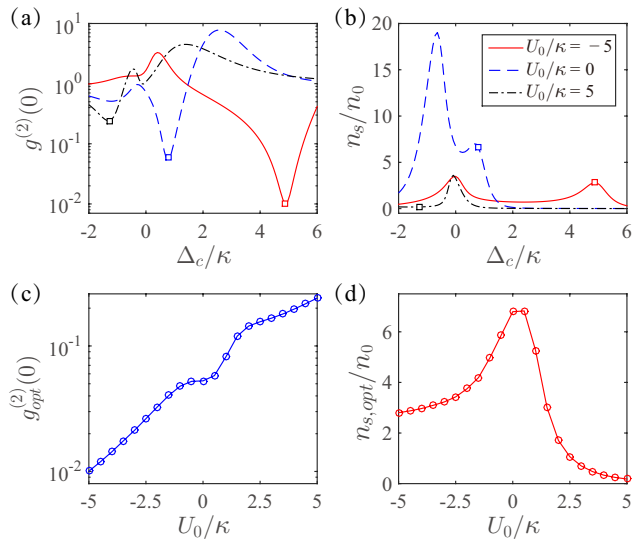}
\caption{(a) and (b) $\Delta_c$ dependence of $g^{(2)}(0)$ and ${n}_s$, respectively, for different $U_0$. The squares mark the position of the minimum values of $g_{\rm{opt}}^{(2)}(0)=\min[g^{(2)}(\Delta_c)]$ for $U_0/\kappa=-5$ (solid line), $U_0/\kappa=0$ (dashed line), and $U_0/\kappa=5$ (dotted-dashed line). Optimal results $g_{\rm{opt}}^{(2)}(0)$ (c) and the corresponding ${n}_{s,\rm{opt}}$ (d) as a function of $U_0$. The other parameter is fixed at $g/\kappa=1$.} \label{main1}
\end{figure}%

More importantly, it is shown that strong PB ($g^{(2)}(0)$) exists in a large parameter region [the blue regime in Fig.~\ref{main}(a)], which is important for experimental feasibility when the system parameters do not perfectly satisfy the optimal conditions. We check that, remarkably, both $g^{(2)}(0)$ and $n_s$ are insensitive to weak dephasing ($\gamma_d/\kappa\ll 1$) in the ground manifold for our system. However, strong dephasing ($\gamma_d/\kappa\gg 1$), e.g., for excited states, will induce significant decoherence in the quantum properties of emitter-cavity coupled systems\cite{Auffeves2010, Majumdar2010, Majumdar2011, Majumdar12, Englund10, Englund12}.

To gain more insight into PB, we plot the $\Delta_c$ dependence of $g^{(2)}(0)$ and ${n}_s$ for different $U_0$ as shown in Fig.~\ref{main1}(a) and (b), respectively. The optical-Stark-shift-mediated red-blue-detuning asymmetry is observed for both $g^{(2)}(0)$ and ${n}_s$. In the absence of $U_0$, an asymmetric structure also appeares, which is ascribed to the quantum interference conditions in the red sideband of the system and has been discussed in detail in our previous work \cite{Tang15}. In contrast to positive $U_0$, which suppresses PB, a negative $U_0$ can significantly enhance PB due to the asymmetric vacuum Rabi splitting of the first manifold in the energy spectrum, where the $U_0$ enhanced anharmonicity of the energy spectrum is conducive to suppressing multiphoton excitations.

Figure~\ref{main1}(c) and \ref{main1}(d) display the minimum value of $g_{\rm{opt}}^{(2)}(0)=\min[g^{(2)}(\Delta_c)]$ and the corresponding ${n}_{s,\rm{opt}}$ at the same value of the cavity-light detuning as a function of $U_0$. As $U_0$ changes from positive to negative, $g_{\rm{opt}}^{(2)}(0)$ decrease approximately linearly. However, the value of ${n}_{s,\rm{opt}}$ displays a peak structure which reaches a maximum  around $U_0=0$. Compared with the case of negative $U_0$, ${n}_{s,\rm{opt}}$ has a rapid decrease with increasing positive $U_0$. Another typical feature is that ${n}_{s,\rm{opt}}$ in the PB regime remains much larger than the empty-cavity photon number when $U_0<0$. The largely enhanced cavity photon number can be readily understood as being due to the additional microwave field. Besides the weak drivening field, the additional cavity field emission ($|0,\downarrow\rangle \leftrightarrow|1,\uparrow\rangle$ transition) originates from Bragg scattering~\cite{Deng14} when the microwave field couples to the $|0,\uparrow\rangle\leftrightarrow|1,\downarrow\rangle$ transition, as shown in Fig.~\ref{path}. Our proposal for realizing a larger cavity output while exhibiting strong antibunching is very important for applications to high-quality single-photon sources~\cite{Birnbaum2005}. Moreover, the $U_0$ dependence of the photon quantum statistics can be realized by a nondestructive diagnosis of the magic wavelength (zero Stark shift), which could facilitate the study of clock transitions in ultracold atomic and ionic gases~\cite{Anders06, Barber08, Goban12, Liu15}.

\begin{figure}[ht]
\includegraphics[width=1\columnwidth]{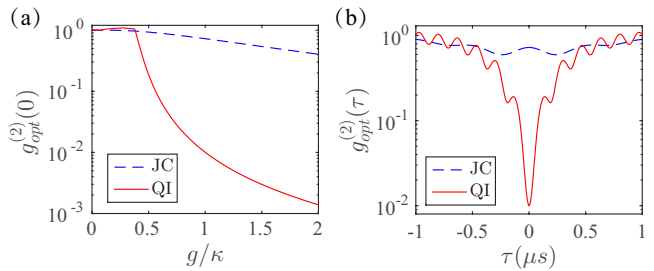}
\caption{(a) $g$ dependence of second-order correlation function $g_{\rm{opt}}^{(2)}(0)$. (b) Dependence of $g_{\rm{opt}}^{(2)}(\tau)$ on the time interval $\tau$ with fixed $g/\kappa=1$. The curves labeled ``JC'' denote the results from the JC model, while ``QI'' denotes the results from our model calculated for the optimal quantum interference conditions for $U_0/g=-5$. }
\label{MR}
\end{figure}

\begin{figure}[ht]
\includegraphics[width=0.9\columnwidth]{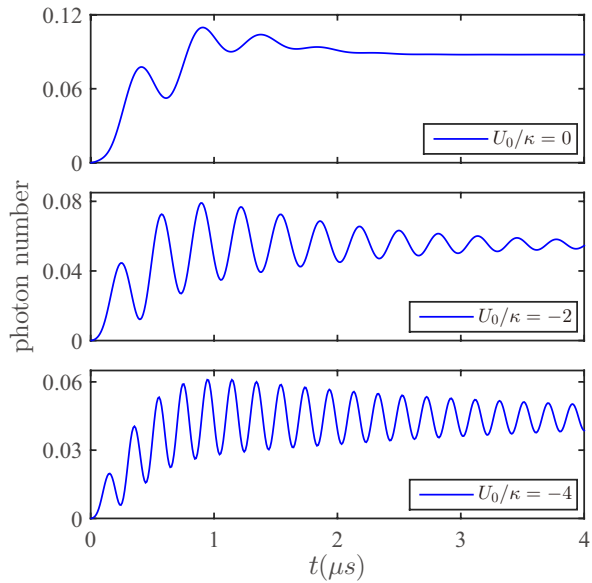}
\caption{Dynamic evolution of the photon number as a function of time $t$ with $g/\kappa=1$. From the first to the third row, the Stark shift is $U_0/\kappa=0$, $-2$, and $-4$.} \label{time}
\end{figure}

We now turn to the validity of the quantum interference mechanism for achieving strong PB in the coupled system. In the absence of the microwave field and the optical Stark shift, the Hamiltonian in Eq.~(\ref{Hamiltonian}) in reduces to the paradigm of the Jaynes-Cummings (JC) Hamiltonian of a two-level atom coupled to a cavity (harmonic oscillator), where strong PB requires a strong atom-cavity coupling $g$ to suppress two-photon excitation~\cite{Birnbaum2005}. Figure~\ref{MR}(a) shows the minimum values of $g_{\rm{opt}}^{(2)}(0)$ for the JC model (dashed line) and our model (solid line) at $U_0/g=-5$ as a function of the atom-cavity coupling $g$. In contrast to the JC model, our $g_{\rm{opt}}^{(2)}(0)$ decreases rapidly when powerful quantum interference conditions are utilized as $g$ increases. Interestingly, strong PB can be easily achieved even when the system is not in the strong-coupling regime ($g/\kappa\sim 1$).
However, the quantum interference mechanism is invalid for a very weak coupling strength due to the finite cavity decay ($g/\kappa<0.38$). In this case, multiphoton ($n\geq3$) excitations begin to play an important role in breaking the PB because of the absence of strong enough energy-spectrum anharmonicity, although the two-photon excitation is eliminated by quantum interference. This explains why the PB of ``QI'' in Fig~\ref{MR}(a) is not enhanced compared with ``JC'' when $g/\kappa<0.38$. Interestingly, we find that the value of the vacuum Rabi splitting $|\Delta_{1,-}|/\kappa$ is equal to 1.97 at $g/\kappa=0.38$, which is equal to the linewidth of the cavity ($2\kappa$). As can be seen, quantum interference can significantly enhance the PB when $|\Delta_{1,-}|/2\kappa>1$ with $g/\kappa>0.38$, which demonstrates that energy-spectrum anharmonicity and quantum interference are two important factors for generating strong PB.

To further confirm the photon statistics, the time-dependent second-order correlation function $g_{\rm{opt}}^{(2)}(\tau)$ is calculated [Fig.~\ref{MR}(b)]. As can be seen, the transmitted photons allow both strong PB, with $g_{\rm{opt}}^{(2)}(0)= 0.01$, and photon antibunching, with $g_{\rm{opt}}^{(2)}(0) < g_{\rm{opt}}^{(2)}(\tau)$. More importantly, $g_{\rm{opt}}^{(2)}(\tau)$ increases to unity at a time $\tau\approx 1/(\kappa+\gamma_e) =1.1$ $\mu $s, which indicates that the nonclassical antibunching effect in the emitted photons possesses a very long coherence time. This lifetime is much larger than that engineered with a single two-level atom (with one ground state and one excited state) trapped in an optical cavity, with a value of around a few tens of nanoseconds~\cite{Birnbaum2005}, which is ascribed to the smaller single-particle loss of the atomic ground state than that of the excited state.

Finally, to illustrate the versatility of the optical Stark shift in our proposal, we plot the time-evolution of the photon number for different values of $U_0$ in Figs.~\ref{time}(a)-\ref{time}(c). We find that the photon number exhibits a quasiperiodic oscillation, with a gradually decreasing amplitude of the oscillation with increasing time. The oscillation behavior of the photon number depicts the quantum interference effect that is used for generating strong PB~\cite{Liew10, Majumdar2012, Vaneph18}. As we can see, the oscillation frequency is very sensitive to the value of $U_0$, which is determined by the Rabi frequency of the system. Interestingly, the photon number rapidly reaches a saturation value for $U_0=0$, as shown in Fig.~\ref{time}(a). In addition, it is clear that a large negative $U_0$ corresponds to a large oscillation frequency due to the enhanced vacuum Rabi frequency [Figs.~\ref{time}(b) and \ref{time}(c)], which can be used to further enhance the understanding of the PB mechanism in our model.\\

\section{EXPERIMENTAL FEASIBILITY}

Our proposal can be readily applied to different single alkali-metal atoms in an optical cavity~\cite{Tobias10, Mucke10}. Taking ${}^{87}$Rb atom as a prototype, we now discuss in detail how to realize our scheme with ultracold $F = 1$ ground states in an experiment. An important ingredient is the realization of a pseudospin-$\frac{1}{2}$ model with the choice $|\uparrow\rangle=|F=1,m_F=0\rangle$ and $|\downarrow\rangle=|F=1,m_F=1\rangle$, where the two desired Zeeman spin states are decoupled from the remaining Zeeman spin state of the $F = 1$ manifold owing to the large quadratic Zeeman shift induced by a bias magnetic field~\cite{Lin2011}. The cavity-mediated Raman coupling could be generated by utilizing the experimental realization of electromagnetically induced transparency with a single atom inside an optical cavity~\cite{Tobias10, Mucke10}. As to the cavity, the single-photon coupling strength and cavity decay rate may be taken as $g_0=2\pi\times 4.5$ MHz and $\kappa=2\pi\times 0.15$ MHz, respectively, where the typical decay rate of an excited atom, $\gamma_e=2\pi\times 3.0$ MHz, is much larger than $\kappa$~\cite{Mucke10, Leonard2017}. The optical Stark shift can be tuned by changing the strength or sign of a large atom-pump detuning, which is a demonstrated capability in current experiments on self-organized superradiance and supersolid phases in an optical cavity~\cite{Mucke10, Mottl, Leonard2017, leonard17}. On the other hand, the optical Stark shift $|U_0|$ is naturally enhanced by further increasing the single-photon coupling strength $g_0$. Importantly, all parameters for generating strong PB in our model are highly controllable in currently available experimental systems. As to the experimental measurement, the cavity transmission can be collected by a single-photon-counting module~\cite{Tobias10}, and the photon quantum statistics can be measured by a Hanbury Brown and Twiss interferometer~\cite{Birnbaum2005}.

\section{CONCLUSION AND DISCUSSION}

In conclusion, we propose a simple experimental scheme to realize strong PB in a single-atom-cavity system by combining a strong vacuum Rabi splitting induced by the optical Stark shift and quantum interference under optimal conditions. We show that the second-order correlation function of the system can be reduced by more than several orders of magnitude compared with the JC model even at a moderate atom-cavity coupling, and, simultaneously, a large cavity photon output can be obtained. As to experimental feasibility, our proposal for generating strong PB can be applied over a fairly large parameter region by tuning the optical Stark shift and the microwave field in a highly controllable way. In particular, in contrast to conventional solid quantum devices, the long coherence time based on the ground-state atom in the system is also extremely conducive to quantum storage and detectors at the single-photon level~\cite{Witthaut2012, Fleischhauer05, Chang14}. Furthermore, the scheme for realizing quantum interference and strong vacuum Rabi splitting beyond the strong-coupling limit could also be applied to explore high-quality multiphoton blockade, e.g., photon-pair sources could be obtained by applying destructive quantum interference to a three-photon excitation~\cite{Chang16}. With slight modifications, our proposal can be extended to various systems, including a single-quantum-dot-cavity model~\cite{Kai2015, Calic11} and a singe-ion-cavity model~\cite{Matthias13, Tian18, Moonjoo19}, which provide several interesting opportunities for exploring the fundamentals of quantum optics, and potential applications in single-photon transistors, all-optical switching, and quantum metrology~\cite{Faraon08, Volz12, Munoz2014}.

\section*{ACKNOWLEDGMENT}\label{acknow}

J.T. is supported by the NSFC (Grants  No. 11804409) and Fundamental Research Funds for the Central Universities (Grant No. 18lgpy80). Y.D. is supported by the NSFC (Grant No. 11874433) and by National Key R$\&$D Program of China (2018YFA0307500). C.H.L. is supported by the NSFC (Grants No. 11874434, No. 11574405).


%

\end{document}